# BIFURCATION AT COMPLEX INSTABILITY

MERCÈ OLLÉ & DANIEL PFENNIGER
*Observatoire de Genève*
*CH-1290 Sauverny, Switzerland*

**Abstract.** The properties of motion close to the transition of a stable family of periodic orbits to complex instability is investigated with two symplectic 4D mappings, natural extensions of the standard mapping. As for the other types of instabilities new families of periodic orbits may bifurcate at the transition; but, more generally, families of *isolated invariant curves* bifurcate, similar to but distinct from a Hopf bifurcation. The evolution of the stable invariant curves and their bifurcations are described.

## 1. Introduction

Complex instability is a type of *generic* instability in Hamiltonian systems with more than 2 degrees of freedom. It was described and named so by Broucke[2] in a study of periodic orbits in the elliptic RTBP, in which he classified all the possible instability types of periodic orbits in 3-degree of freedom Hamiltonian systems. In real problems, it appears in celestial mechanics[5] and galactic dynamics[3],[4],[8],[10],[11],[12], in particle accelerator problems[7], and actually, in many engineering problem with 3 or more degrees of freedom. On the other hand, complex instability has been studied by mathematical techniques only during about the last decade[1],[6],[9].

In this paper, we are concerned with the intricacies in phase space associated to the transition from stability to complex instability. We refer to the numerical study of this transition as well as the description of the bifurcating objects done by Pfenniger[10]. We shall briefly review those results and we shall concentrate, specially, on the evolution of the *stable bifurcating invariant curves* when varying some parameter. There appear two types of bifurcations resembling the bifurcations of periodic orbits: the 'period-doubling' bifurcation and the bifurcation of a pair of asymmetric invariant curves.

## 2. Symplectic mappings

As well known, Hamiltonian systems can be transformed into symplectic mappings by means of the Poincaré sections. In 3-degree of freedom systems, there corresponds a 4D section space. Thus, let us consider a symplectic mapping $T$, 0 a fixed point, $A$ the Jacobian matrix around the origin, and $\lambda$, $1/\lambda$, $\mu$, and $1/\mu$ the eigenvalues of $A$.

From stability, there are only three routes to instability as a system parameter is changed: 1) A pair of eigenvalues $\lambda$, $\bar{\lambda}$ on the unit circle in the complex plane bifurcates onto the real axis at $(1,0)$ –tangent bifurcation–; 2) the same occurring at $(-1,0)$ –period doubling bifurcation–; 3) the four eigenvalues collide simultaneously by conjugate pairs on the unit circle and leave it into the complex plane; this is the transition to complex instability, also called Krein collision. Let us denote $k$ the rotation number corresponding to the ratio between the modulus of the eigenvalues and $2\pi$. If $k$ is rational, $k = p/q$, where $q$ is the smallest positive integer, then $q$ *periodic orbits* may bifurcate, because the eigenvalues are then integer roots of 1. But, usually, $k$ is irrational and there are no bifurcating families of periodic orbits but of invariant curves at the transition point.

In order to describe the transition from stability to complex instability we have used[10] two 4D natural generalizations of the standard map inspired by the symplectic Froeschlé map, the mappings $T_s$ and $T_t$,

$$T_s \begin{pmatrix} x_1 \\ x_2 \\ x_3 \\ x_4 \end{pmatrix} = \begin{pmatrix} D[x_1 + K_1 \sin(x_1+x_2) + L_1 \sin(x_1+x_2+x_3+x_4)] \\ x_1 + x_2 \\ D[x_3 + K_2 \sin(x_3+x_4) + L_2 \sin(x_1+x_2+x_3+x_4)] \\ x_3 + x_4 \end{pmatrix} \pmod{2\pi} \quad (1)$$

$$T_t \begin{pmatrix} x_1 \\ x_2 \\ x_3 \\ x_4 \end{pmatrix} = \begin{pmatrix} D[x_1 + K_1 \sin(x_1+x_2) + L_1 \tan(x_1+x_2+x_3+x_4)] \\ x_1 + x_2 \\ D[x_3 + K_2 \sin(x_3+x_4) + L_2 \tan(x_1+x_2+x_3+x_4)] \\ x_3 + x_4 \end{pmatrix} \pmod{2\pi} \quad (2)$$

The dissipation parameter $D$ is only used for computing purposes. When $D < 1$, $D = 1$, $D > 1$ the mappings are respectively volume contracting, preserving and dilating. We study the mappings around $\mathbf{x} = 0$; they have the same Jacobian $A$ but different non-linear properties as described below. We restrict the parameter space by taking $L_1 = -L_2 \equiv L$, $K_1 \equiv K$ and $K_2 = 0$; the transition takes place when $L = -K/4 \equiv L_{crit}$ in the interval $-8 < K < 0$. $L$ is a varying parameter and $K$ assumed not to depend on $L$ (see [10] for a discussion of the parameters).

## 3. Transition stability-complex instability

Striking phase structures appear at the transition from stability to complex instability. The main results are that, as in classical pitchfork bifurcations, the bifurcating structures may be "direct" (the bifurcating objects unfold on the unstable side), or they may be "inverse" (the bifurcating objects unfold on the stable side) as the parameter $L$ is varied. For the mapping $T_s$ ($T_t$), the bifurcation is direct (inverse), and the bifurcating objects exist for $L > L_{crit}$ ($L < L_{crit}$).

When the rotation number $k$ is rational, then we find $q$-periodic orbits which bifurcate. In [10], a detailed description of such periodic orbits and their stability properties are given.

When the rotation number $k$ is irrational, *stable* (*unstable*) invariant curves bifurcate for the mapping $T_s$ ($T_t$). In order to find the stable invariant curves for the mapping $T_s$, we exploit the property that they become attracting limit cycles as soon as dissipation ($D < 1$) occurs. Starting at $D < 1$, the method consists in progressively increasing $D$ while controlling the convergence of the consequents onto the invariant curves. When $D = 1$, we have reached the desired invariant curve. Fig. 1 shows such an example of bifurcation.

We are now interested in describing the typical bifurcations along of a family of stable invariant curves when varying some parameter ($L$). Let us fix a value of $K$ such that we have an irrational $k$ and let us study the family of stable invariant curves of $T_s$ which bifurcates at the transition. In order to describe it, we define, for a given invariant curve, $x_{20}$ the value of $x_2 > 0$ when "$x_1 \approx 0$", that is, when $||x_1|| \le \epsilon$, $\epsilon$ a small positive quantity.

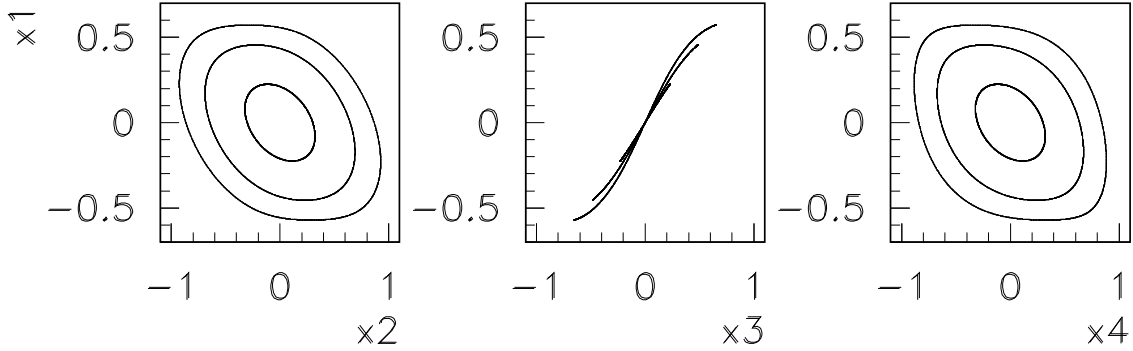

*Figure 1.* Projections of the invariant curves in mapping $T_s$, $K = -1$, for values $L = 0.26$ (inner one), $L = 0.30$, $L = 0.35$ (outer one) ($L_{crit} = 0.25$)

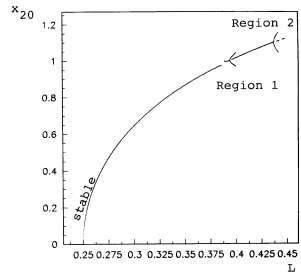

*Figure 2.* Evolution of the stable bifurcating invariant curves for mapping $T_s$, $K = -1$, $L_{crit} = 0.25$

At first, the invariant curves are stable, but as $L$ increases, the stability character changes and there appear other bifurcations (see Fig. 2, where a representation of the invariant curves by their $x_{20}$ value when varying $L$ is given). In particular, we distinguish two different regions in Fig. 2, corresponding to two different kinds of bifurcations. We describe region 1 (Fig. 3) with some detail: we notice, first, a deviation of the family of stable invariant curves, for $L > 0.3805$; there appear slightly unstable invariant curves. We also remark a gap between $L = 0.387$ and $L = 0.388$, which is due to a resonance associated to a family of 28-periodic orbits, which has a transition from stability to complex instability; thus, we may expect new bifurcating families of invariant curves there, which we have not followed yet. Finally, there is a change of stability (by a pitchfork bifurcation) and there bifurcate "double-periodic" invariant curves (the invariant curve makes 2 turns before closing in section space) apart from the central invariant curve which becomes unstable. Fig. 4 shows a 'double-periodic' invariant curve. Some branches of invariant curves in this figure remain to be continued; presumably they become unstable and our numerical method is unable to determine unstable invariant curves.

We consider now region 2 (Fig. 2). Again a change of stability occurs together with a new pitchfork bifurcation. It has, however, a different meaning here: for a fixed value of $L$, $L > 0.435$, we have three values of $x_{20}$, the external ones which correspond to *two different* asymmetric stable invariant curves and the central one which is the central invariant curve which has become unstable. We show two asymmetric invariant curves in Fig. 5.

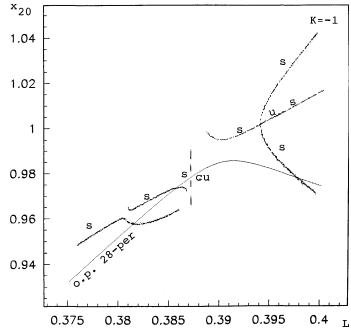

*Figure 3.* Bifurcations of the invariant curves in region 1

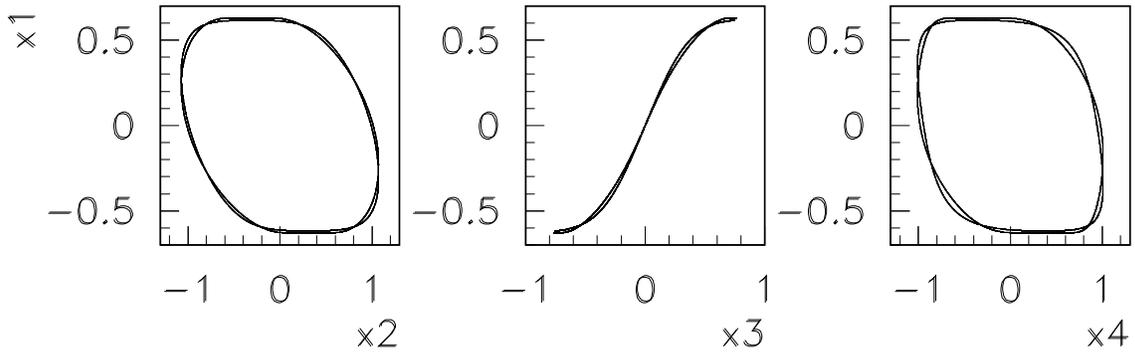

*Figure 4.* 'Double-periodic' invariant curve in Region 2, mapping $T_s$, $K = -1$, $L = 0.395$

The effect of the bifurcation on the chaotic neighbourhood of the central periodic family can be summarized as follows.

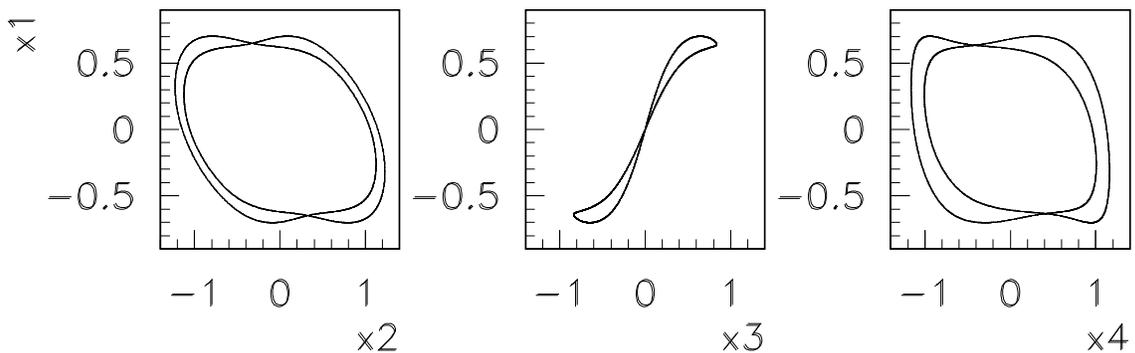

*Figure 5.* Two asymmetric bifurcated invariant curves; mapping $T_s$, $K = -1$, $L = 0.44$

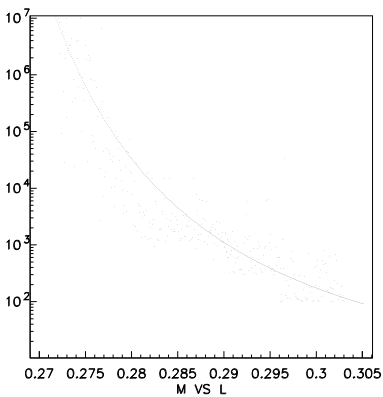

*Figure 6.* Nekhoroshev estimation for $m$

In the direct case (mapping $T_s$), the chaotic orbits originating around the central complex-unstable orbit are *confined* for a long time. As an example and in order to give a Nekhoroshev estimation of the powerful confinement, we fix a number of iterates $N$ (say $N = 10^3$), we fix the initial conditions close to the origin (say $x_i = 0.01$, for $i = 1, 2, 3, 4$), we define $d = \max_{n \leq N} ||T_s^n(\mathbf{x})||$, $\mathbf{x} = (x_1, x_2, x_3, x_4)$ and we compute $d$ when varying the parameter $L$, for a $K$ fixed. Afterwards, we compute the necessary number $m$ of iterates in order to have, for each value of $L$, a value of $d$, $d'$, such that $d' \geq 1.01d$. For $K = -1$, we observe a huge number of iterates $m$ between $L_{crit} = 0.25 < L < 0.27$, and we can fit $m$ by a Nekhoroshev expression $n \approx \exp[0.08418/(L - L_{crit})^{1.37385}]$, for $0.27 \leq L \leq 0.305$ (see Fig. 6). To summarize, invariant curves and *confined* chaotic orbits coexist.

In the inverse case (mapping $T_t$), the transition to chaos is immediate and no confinement occurs. The invariant tori surrounding the stable central family dissolve before the transition to complex instability. See [10] for the numerical details to reach the last invariant torus by means of the dissipation parameter.

**Acknowledgments** M. Ollé thanks Prof. G. Gómez for many useful discussions, and Dr. P. Comas for his help in all computer and software issues. M. Ollé is partially supported by DGICYT grants PB90-0580 and PR95-071, and D. Pfenniger by the Swiss FNRS.

M. Ollé's usual address: E.T.S.E.I.B., Dept. Matemàtica Aplicada I, Diagonal 647, 08028 Barcelona, Spain.